# Two Phases Leakage Detection Strategy Supported by DMAs

G. Messa[1], G. Acconciaioco[2], S. Ripani[3], L. Bozzelli[4], A. Simone[5], O. Giustolisi[6]*

[1] Acquedotto Pugliese S.p.A., via Cognetti, 36, Bari, Italy
[2] Politecnico di Bari, via Orabona, 4, Bari, Ital
[3] Università degli Studi "G. D'Annunzio", Viale Pindaro 42, Pescara, Italy
[4] Politecnico di Bari, via Orabona, 4, Bari, Italy
[5] Politecnico di Bari, via Orabona, 4, Bari, Italy
[6] Politecnico di Bari, via Orabona, 4, Bari, Italy

*Corresponding author. Tel: +39-329-317-094. E-mail: *orazio.giustolisi@poliba.it*

**Abstract**

The present work proposes a novel two phases *model-based* strategy for leakage detection. The two phases are: the identification of the district metering area (DMA) and the pipe pre-localization into the identified DMA. The strategy is based on detecting and pre-localizing the punctual leakage as anomaly with respect to the normal working conditions.

A further novelty is the fact that the pre-localization phase returns the sequence of pipes to inspect, which makes the strategy attractive for water utilities, whose aim is to identify the anomaly at DMA level and, successively, to localize it with the minimum inspection cost.

Furthermore, a *random database* is useful to test the performance of the strategy with respect to the configuration of DMAs and the pressure metering system. Consequently, a novel strategy to design the location of pressure meters is also proposed. It is demonstrated that the entire strategy limits false positives during the DMA identification phase by using the recently proposed index named Asset Management Support Indicator (AMSI). AMSI is invariant with respect to the deterioration, i.e., it is sensitive to its increase causing punctual leakage.

The strategy is studied and discussed using two real Apulian WDNs managed by Acquedotto Pugliese.

## 1. Introduction

Real water losses represent one of the most critical challenges in the management of urban water distribution networks (WDNs). Losses are an indicator of the system "health" as they are the indicator of the deterioration and pressure management (Giustolisi et al., 2024). The impact of water loss management is relevant for socio-economic sustainability, i.e. next generations, because of the huge value of that asset. Furthermore, the water loss management can impact on environmental sustainability, reliability, affordability, etc. of the WDNs and more in general water supply systems.

Real water losses are outflows from pipes, joints and property connections, resulting from the natural deterioration of materials and pressure constraints for service requirements. In fact, at the system scale, they depend on system deterioration and pressure status according to Torricelli's law (Giustolisi et al., 2024).

The deterioration is due to the aging of pipes and the history of pressure management, mainly caused over the years by external factors, such as operating pressure variations, transient hydraulic events, mechanical stresses and temperature fluctuations (Kleiner et al., 2001; Kleiner et al., 2002).

At the local scale, real water losses are generated by single outflows, which are characterized by specific outflow coefficients and local pressures, according to Torricelli's law as well. However, their spatial distribution is unknown, and the outflow coefficients (related to local deterioration) and pressures can be highly variable, consequently the outflows can be variable as well.

Therefore, being leakages a population of local effects that can be distinguished based on their detectability, the scientific-technical literature classifies them into categories (Lambert, 1994): the diffuse leakages, also named *background leakages* and the punctual leakages, also named *bursts* if we refer to the active control.

Note that diffuse leakages evolve over time into punctual leakages because of internal and external local factors (Kleiner et al., 2001; Kleiner et al., 2002). A relevant challenge for water loss management is, then, to promptly detect their evolution, both for water loss reduction and system reliability related to service interruptions.

For this reason, the technical scientific literature, especially of recent years, reports many contributions about punctual leakage management (Puust et al., 2010; Hu et al., 2021; Rajabi et al., 2023a; Wu et al., 2024) i.e. detection, pre-localization and localization, which are mainly based on detecting the generation of an anomaly related to the acoustic effect of the cavitation of the outflow, the local pressure drops, the water balance at District Metering Areas (DMAs), etc.

For the purpose of the literature review, punctual leakage detection strategies can be divided into *field-based* and *model-based* strategies, which use, in different manners, measurements, whose collection is made easier by the ongoing digital transition (Giustolisi et al., 2024; Giustolisi, 2023).

*Field-based* strategy is generally related to instrumental techniques including acoustic and non-acoustic detection. Acoustic detection relies on capturing the signals generated by leakages in pressurized pipelines due to cavitation phenomena related to outflow from holes. Few examples are listening-rods, noise loggers (Puust et al., 2010), correlators, and pig-mounted acoustic sensors (Mergelas et al., 2005).

Alternative non-acoustic techniques have been also developed, offering good solutions when background noise or soil conditions hinder acoustic signal detection. Few examples are gas injection, ground penetrating radar and infrared photography (Li et al., 2015).

Note that *field-based* strategy is expensive because of equipment acquisition, time-consuming of inspections and operators involved. Therefore, the integration of *field* and *model-based* strategies can reduce costs (Puust et al., 2010).

*Model-based* strategy infers anomalies comparing measurements (typically pressure and flow data) with model-predicted ones considering the normal working conditions of the hydraulic system. They can be divided into two main categories: transient and steady state analysis methods.

About transient analysis, Liggett and Chen (Liggett et al., 1994) formulated leak detection as an inverse optimization problem, aimed at minimizing the difference between measured and simulated transient pressure responses generated by a punctual leakage. This approach was further enhanced by exploring optimization techniques to improve or accelerate the solution of the inverse problem. For instance, Vitkovsky et al. (Vítkovský et al., 1999; Vítkovský et al., 2000) explored the use of Genetic Algorithms (GAs) and Kapelan et al. (Kapelan et al., 2003) introduced a hybrid genetic algorithm (HGA) combining GAs with the Levenberg–Marquardt method.

Alternative approaches based on time and frequency domain analyses were also proposed in (Brunone et al., 2001; Mpesha et al., 2001). Initial studies employed Fast Fourier Transform technique to extract frequency-domain signatures associated with leakages (Jonsson et al., 1992; Wang et al., 2002;. Subsequently, Ferrante et al. (Ferrante et al., 2009) proposed the use of wavelet transform to retain information coming from the time domain analysis. This method was further improved by subsequent studies (Brunone et al., 2022).

In recent years, steady state analysis has been widely adopted in leakage detection for water distribution networks. This approach requires a good calibration of the mathematical model to obtain reliable results (Sumer et al., 2009; Savic et al., 2009) but also advanced hydraulic analysis to obtain a phenomenological twin inside the concept of digital twin to obtain a predictive model (Giustolisi, 2023).

Wu and Sage (Wu et al., 2008) proposed a model calibration approach employing Genetic Algorithms (GAs) to optimize both the localization of leakages and their outflow coefficients, minimizing the differences between measured and simulated pressure and flow data.

Similarly, Sanz (Sanz et al., 2016) integrated leakages detection into the calibration process by comparing calibrated parameters to historical values to distinguish between system evolution and leakage. Then, the geographical distribution of demand parameters allows leak localizations.

The role of sensor placement in supporting such methods has also been explored. The effectiveness of GAs and multi-objective optimization techniques in identifying optimal pressure sensors placement was studied in several works (Kapelan et al., 2005; Ferreira et al., 2022).

Other *model-based* approaches rely on sensitivity analysis. First introduced by Pudar and Ligget (Pudar et al.,1992), the simulation is conducted under both normal and leakage conditions to estimate the pressure sensitivity at each node.

Successively, Pérez (Pérez et al., 2011) proposed an optimal sensor placement methodology by combining a sensitivity matrix, binarized using a threshold, and genetic algorithms, to maximize leakages detectability.

Currently, driven by technological advancements and increasing computational power, there has been a growing interest in incorporating machine learning techniques to enhance and accelerate leakage detection processes. Some algorithms, i.e. Statistical Classifiers (Corzo et al., 2023), Bayesian Classifiers (Soldevila et al., 2017), Support Vector Machines (Zhang et al., 2016), Convolutional Neural Networks (Javadiha et al., 2019; Xu et al., 2025), Random Forests (Fereidooni et al., 2021), and Multiscale Fully Convolutional Networks (Hu et al., 2021), have been used to train classifiers capable of identifying and localizing leaks. However, the number of punctual leakage events, required to train such overparameterized approaches, limits practical applications.

The present work proposes an innovative two phases strategy for the identification and pre-localization of unreported punctual leakages supported by DMAs using advanced hydraulic analysis, i.e. steady state. The first phase identifies the DMA where the evolving punctual leakage occurs, analysing a deterioration factor whose value increases with the evolution of the punctual leakage size (Giustolisi et al., 2024); while the second phase pre-localizes the punctual leakage, analysing the pressure anomaly into the identified DMA and returning the sequence of pipes to inspect to repair it.

The DMA identification phase uses the recently proposed Asset Management Support Indicator (AMSI), which is invariant with respect to the deterioration, i.e. it is sensitive to its increase causing a punctual leakage. The two-phase strategy limits false positives while also accounting for pressure measurement errors.

The strategy requires a phenomenological twin of the WDN that realistically replicates the network's structure and hydraulic behaviour, increasing the predictive capabilities with respect to standard hydraulic modelling methods (Giustolisi et al., 2008; Giustolisi et al., 2024).

Finally, the proposed strategy integrates testing of the pressure metering system supporting the sampling design of pressure meters.

The next paragraph expands and explains the technical-scientific innovations proposed in this work; the third paragraph describes the methodology underlying the two phases strategy; the fourth paragraph explains and clarifies the necessity of the advanced hydraulic analysis and AMSI indicator for the successful implementation of the strategy; the fifth paragraph illustrates the strategy for pressure sampling design; the sixth paragraph reports a case study: at first, the analysis is performed in *error-free* condition and successively, considering *pressure measurement errors* and a *threshold for AMSI variation*.

## 2. Novelties and motivations

The aim of this paragraph is to discuss, from a scientific and technical standpoints, the effectiveness of the proposed novelties for leakage detection.

The main novelties can be summarized as follows:

- a *two phases strategy* involving DMAs for anomaly detection (first phase) and pipe pre-localization (second phase).
- The *pre-localization of a sequence of pipes* to inspect instead of a single pipe.
- The *progressive integration of field-data* to enhance predictive capabilities.
- The use of a *phenomenological twin* involving the advanced hydraulic analysis and simulation of the punctual leakage into pipes instead of network nodes.
- The use of *offline scenario database* to generate punctual leakage events varying the boundary conditions.
- The use of *offline random database* to assess the performance of the leakage detection with respect to the identification phase (DMAs) and the pressure metering system (pre-localization).
- The development of a novel *strategy to design the pressure metering system* considering the system engineering supported by DMAs.
- The possibility of considering *metering errors* during leakage detection and assessing the performance.

The *two phases strategy* is supported by the metering systems of districtualization, which allows the identification of the DMAs where the leakage occurs before inspecting. This is one of the most important motivations of system engineering by means of DMAs, which also justifies the related investments. The real advantage is the reduction of the network area to inspect and the related costs, being the pre-localization phase performed on a mean network length equal to the mean length of DMAs on average. Furthermore, the use of the novel Asset Management Support Indicator (AMSI) (Giustolisi et al., 2024), based on density indicator of water losses $D_{leak}$, computed by means of DMA meters, and the mean pressure status, allows reducing the false positive. Note that AMSI is a deterioration factor increasing with the evolution of the punctual leakage size (Giustolisi et al., 2024), therefore it is effective to identify the DMA where anomaly is evolving.

The second phase, the *pre-localization of the sequence of pipes* to inspect, prioritizes the pipes and, if the false positive is avoided in the first phase, it is a matter of operational efficiency and can be improved with field information, historical data, operator knowledge, etc.

The strategy supports the *progressive integration of field-data* to enhance predictive capabilities. In fact, the *prior information* on the leakage propensity of each pipe can be integrated into the design of the *sequence of pipes* to inspect. For instance, asset information like pipe age, material, diameter and connections to properties etc., or field information, like historical failures, pipe repairs, noise loggers, correlators, etc., can be progressively part of the design of the *sequence of pipes* to inspect enhancing leakage detection over time. Note that the reported strategy is appealing for water utilities because it integrates information and data, creating value from different management activities.

Being the proposed strategy *model-based*, the hydraulic analysis is crucial for predictability of leakage scenarios. To this purpose, the phenomenological twin concept (Giustolisi, 2023) is fundamental because it integrates diffuse leakages, which are modelled at pipe scale depending on pressure and pipe deterioration (Giustolisi et al., 2024). In fact, diffuse leakages analysis is mandatory for model predictability because the increase of the outflow of a punctual leakage reduces the local diffuse leakages which are a relevant boundary condition of the hydraulic state of the system. Furthermore, the pressure-driven analysis of the *phenomenological twin* is mandatory when a punctual leakage causes pressure deficient conditions for consumers (Giustolisi et al., 2012), or the system is characterized by private tanks. Additionally, the *phenomenological twin* concept supports the enhancement of the geometric representation of the system. For example, the consumers are not concentrated in nodes and their representation in the real georeferenced positions is consistent with smart metering and allows using those consumption data, real elevations, etc. in line with digital transformation and, again, with model predictability. The punctual leakage into pipes instead of network nodes is, then, crucial for pipe pre-localization.

The use of *offline scenario database* creates the possibility to consider different boundary conditions as reference to the detection and pre-localization phases avoiding time-consuming optimization or analyses, which need to be performed every day for many consumptions centres.

The use of *offline random database* to assess the performance is crucial in *model-based* leakage detection because the variable complexity of WDNs, configuration of DMAs, density of pressure metering, ask for the assessment of the real capacity of the strategy.

In fact, the development of a novel strategy to design the pressure metering system is based on the use of an *offline random database*.

Finally, the *offline random database* can be integrated with a generation of measurement errors to assess the reduction of the performance with the decrease of the measurement accuracy.

The following paragraph presents the detailed methodology of the proposed *model-based* leakage detection.

## 3. Methodology

Figure 1 illustrates the *two phases strategy*, identification and pre-localization of punctual leakages. To the purpose of generating effective *scenario* and *random databases*, the predictive modelling capability is fundamental. The *phenomenological twin* (Giustolisi, 2023) is then mandatory as discussed in the previous paragraph. In fact, advanced hydraulic analysis is used to reproduce hydraulic behaviour under normal function conditions, e.g., varying consumer demand and, consequently, pressure-driven diffuse leakages (Giustolisi et al., 2024). Furthermore, advanced hydraulic analysis is essential to predict the perturbed operating conditions once a punctual leakage is assumed into a pipe to build the databases. In fact, both the databases are generated by analysing scenarios, which are built sampling the anomaly generated by a leakage in each pipe with respect to the normal condition, i.e. the same boundary condition without the pipe leakage.

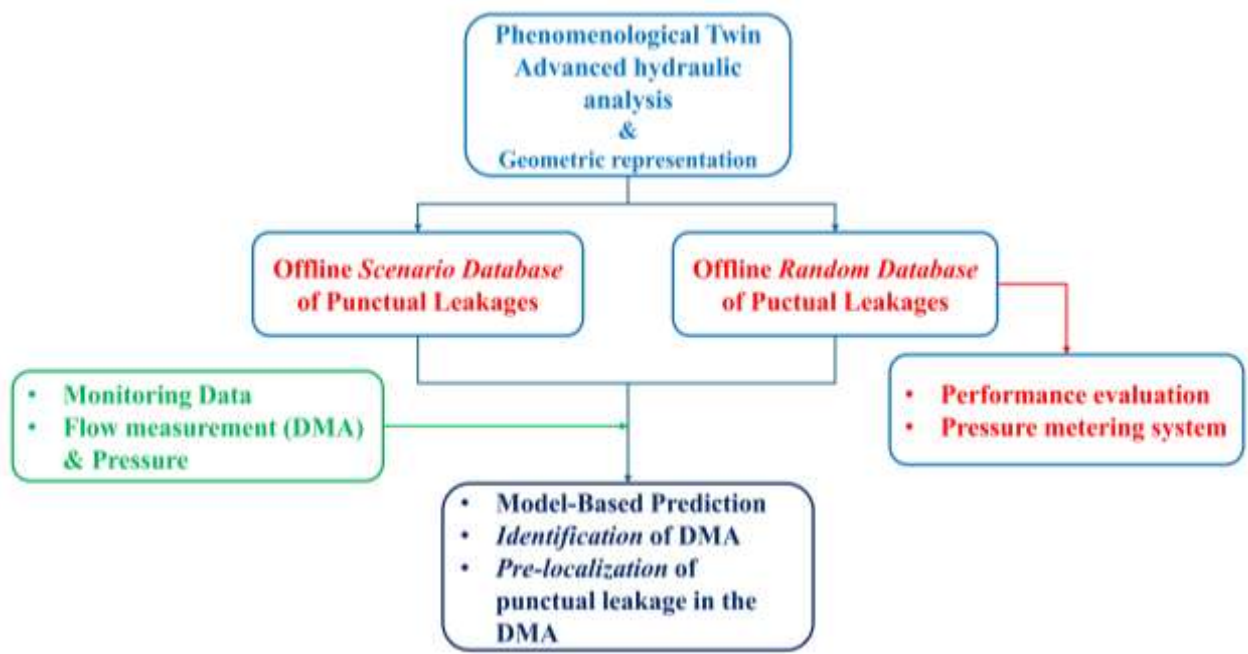

Figure 1. Flowchart of the leakage detection strategy

Note that, as previously reported, the *phenomenological twin* relies on the advanced geometric representation of the hydraulic system, i.e., its domain, the network, being complex network theory integrated (Ciliberti et al., 2023). This means that the analysis of DMAs is realistic because the boundary closed valves are pipes disconnected in the network topology. Thus, the hydraulic analysis is predictive and any kind of analysis at the DMA level is effectively supported by the *phenomenological twin* (Laucelli et al., 2023).

In addition, the complex network theory also supports the representation of punctual leakage which is integrated into the *phenomenological twin*, and whose representation is shown in Figure 2.

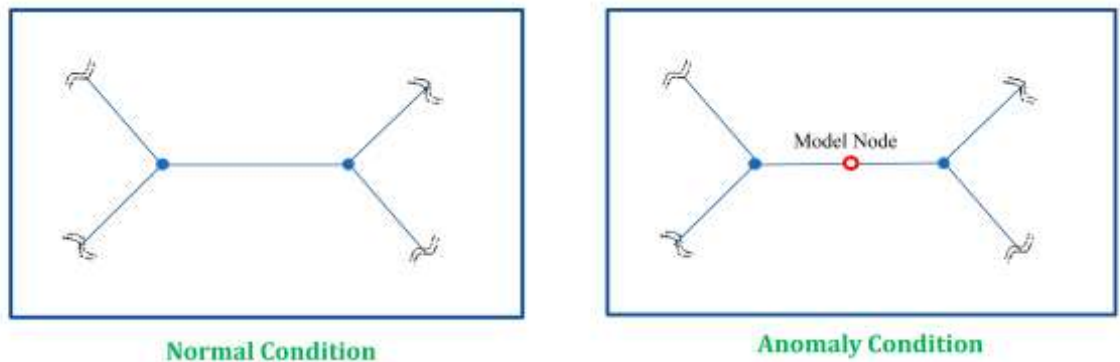

Figure 2. Dynamic modification of WDN topology in the presence of punctual leakage

The representation of the punctual leakage along the pipes, for example in their midpoint, is mandatory to generate the two databases by sampling the abnormal conditions due a leakage in each pipe varying its size. This way, it preserves the information about the sequence of the pipes to be inspect, as reported in the previous paragraph, that otherwise results lost if the leakages are represented in the nodes of the network.

The *scenario database* is then built assuming for each pipe of the network a leakage. This is performed varying its size assuming a circular orifice. The diameter of the orifice has a minimum and maximum values, which are driven by the characteristics of the specific hydraulic system.

It is necessary to remind that the two phases strategy involves the detectability at the DMA level, meaning that the leakage outflows depend on the orifice size and system pressure.

Therefore, the minimum value represents a threshold about the identifiability at DMA level related to the variation network flow status, i.e., to AMSI deterioration index. The maximum value represents an upper bound of the range for which the identifiability at DMA level becomes trivial because of the high variation of the flow status, i.e. the identification of the variation of AMSI deterioration index is trivial.

Differently, the capability of pre-localization into the identified DMA and the accuracy of the pipe sequence to inspect mainly depend on the variation of the pressure status, which is driven by the hydraulic capacity of the specific system. E.g., a high leakage outflow can allow detectability without a good accuracy of the sequence of the pipes to inspect; for this reason, it is important to integrate other information as reported in the previous paragraph.

An open issue is the number of orifice sizes to be into the range; it is a matter of database dimension and runtime to build it. It is experienced that, starting from five orifice sizes, the leakage detection strategy works, and further increasing their number does not lead to a significant improvement in accuracy.

About the varying boundary conditions of the *scenario database*, they refer to the normal ones.

Varying the size in the assumed range, the *scenario database* is computed for different levels of consumer consumption considering the minimum, maximum and mean values available from the operative cycle data of the specific hydraulic system.

It is worth nothing that the system works considering the average daily status of WDN, more realistic for water utilities to scheduling intervention. Thus, the monitoring values are more robust with respect to the single measurement errors which are influenced by the natural dynamic behaviour of the system not modelled in steady-state analysis.

Night-time conditions are more suitable for DMA identification due to higher diffuse leakages and pressures, which enhance the robustness of AMSI variation calculations. Conversely, day-time conditions are preferred for pre-localization as the higher level of consumption make the hydraulic capacity lower, resulting in increased pressure variations.

The *random database* is built like to the *scenario* one, but the size of the orifices and the consumer demand level are randomly generated using a uniform probability distribution in the same range. The pipes where the leakage occurs are also randomly selected.

A minimum size of the *random database* is useful for a robust assessment of the leakage detection strategy performance from a statistical standpoint.

The leakage detection with real data needs of flow meters and pressure at DMAs. The flow meters allow calculating the diffuse leakages, i.e. the density indicator of water losses $D_{leak}$ for AMSI.

The ongoing smart metering at the consumers due to the digital transition is a perspective of increasing accuracy. Without smart metering, the advanced hydraulic analysis involving the leakage model can help in determining the level of consumption for the specific operative cycle (Laucelli et al., 2023; Giustolisi et al., 2024).

AMSI is compared with the same value of the normal conditions and being a deterioration index, it varies if a leakage size is evolving into the DMA (Giustolisi et al., 2024). The maximum AMSI identifies the DMA with the anomaly while the threshold is useful for a reliable identification considering errors and uncertainties. If the AMSI variation does not exceed the defined threshold, the punctual leakage cannot be identified, as it cannot be reliably associated with any specific DMA. In other words, the threshold of the variation of AMSI is useful to focus on the most relevant punctual leakage configurations, thereby avoiding false positives in DMA identification.

The strategy, then, can identify more than one DMA and it is a matter of operational management to decide where to start inspection.

In addition, the identification strategy does not depend on the number of leakages evolving into a DMA because AMSI is a general deterioration factor for it.

Once identified the DMA, the pre-localization phase involves the variation of pressure data of that DMA with respect to the normal condition. Then, the sequence of pipes to inspect is determined using the Pearson's linear correlation between those data and the *scenarios database* resized to the pipes of the identified DMA. Accounting for measurement errors the correlation might be performed adopting a conservative strategy of using meters whose variation is higher than the possible accuracy error user-defined. The use of alternative correlation techniques, e.g., based on machine learning, is out of the focus of this work presenting a general strategy. It could be explored in future works.

The *random database* works as generator of leakage events to test the strategy with respect to errors and pressure meter sampling.

## 4. Advanced hydraulic analysis and AMSI

The proposed strategy requires the use of a *phenomenological twin*, based on the advanced hydraulic analysis, which integrates diffuse leakages at pipe level depending on pressure and pipe deterioration (Giustolisi et al., 2024). Furthermore, advanced hydraulic analysis provides significant advantages over traditional methods because it supports the predictivity of the model, also with respect to burst scenarios, which are abnormal functioning conditions, being phenomenological.

The mathematical model for steady-state advanced hydraulic analysis of a pressurized water pipeline system, composed of $n_p$ pipes with unknown flow rates, $n_n$ nodes with unknown heads and $n_0$ nodes with known heads, is expressed by two sets of momentum and mass balance equations at pipes and nodes, respectively, as follows (Giustolisi et al., 2024):

$$\begin{aligned}
\mathbf{A}_{pp}\mathbf{Q}_p^{mean}(t,\Delta t) + \mathbf{A}_{pn}\mathbf{H}_n^{mean}(t,\Delta t) &= -\mathbf{A}_{p0}\mathbf{H}_0^{mean}(t,\Delta t) \\
\mathbf{A}_{np}\mathbf{Q}_p^{mean}(t,\Delta t) - \mathbf{d}_n^{mean}\left(t,\mathbf{H}_n^{mean}(t,\Delta t)\right) &= \mathbf{0}_n \\
\mathbf{d}_n^{mean}\left(\mathbf{H}_n^{mean}(t,\Delta t)\right) &= \frac{\int_t^{t+\Delta t}\mathbf{d}_n\left(\mathbf{H}_n(t,\Delta t)\right)}{\Delta t}
\end{aligned} \quad (1)$$

where $\mathbf{A}_{pp}=[n_p, n_p]$ is a diagonal matrix with elements based on the pipes resistance; $\mathbf{Q}_p^{mean}=[n_p, 1]$ is a column vector of unknown pipes flow rates; $\mathbf{H}_n^{mean}=[n_n, 1]$ is a column vector of unknown nodal heads; $\mathbf{H}_0^{mean}=[n_0, 1]$ is a column vector of known nodal heads; $\mathbf{0}_n=[n_n, 1]$ is a column vector of null values; $\mathbf{d}_n^{mean}=[n_n, 1]$ is a column vector of nodal water flows; $\mathbf{A}_{pn}=\mathbf{A}_{np}^T$ and $\mathbf{A}_{p0}$ are the topological incidence sub-matrices of size $[n_p, n_n]$ and $[n_p, n_0]$, respectively, derived from the general topological matrix $\bar{\mathbf{A}}_{pn}=[\mathbf{A}_{pn} \mid \mathbf{A}_{p0}]$ of size $[n_p, n_n+n_0]$. Note that the superscript *mean* refers to the fact that the model is steady state in $(t, t+\Delta t)$, where $\Delta t$ is the hydraulic timestep, and, therefore, the status variables and boundary conditions are the mean values over the $\Delta t$, as well described in (Giustolisi et al., 2024).

The hydraulic model in (1) is the general form for pressure-driven analysis (PDA) (Giustolisi et al., 2012) meaning that the hydraulic status of the water system, $\mathbf{Q}_p^{mean}$ and $\mathbf{H}_n^{mean}$, is driven by pressure dependency of nodal demands, $\mathbf{d}_n^{mean}$. This is a more general assumption than demand-driven analysis (DDA) assuming the priors about $\mathbf{d}_n^{mean}$. Note that $\mathbf{d}_n^{mean}$ represents the summation of different components of demands, as for example customer and leakage ones (Giustolisi et al., 2012).

The advanced hydraulic model supports pressure-deficit scenarios with respect to consumers and pressure-dependent diffuse leakages at the single-pipe level (Giustolisi et al., 2024; Giustolisi et al., 2008). Note that burst scenarios modify the pressure status of the hydraulic system and it results in change of the diffuse leakages, which include the punctual leakages, until they are not detected, localized and repaired. Diffuse leakages for each $k$th pipe are computed using the empiric Germanopoulos' POWER model (Germanopoulos, 1985) or the experimental and physically based FAVAD one (May, 1994; Van Zyl et al., 2014) as following reported (Giustolisi et al., 2024):

$$d_{k-leak}(t, \Delta t) = \begin{cases} \beta_k^{\text{Power}} \cdot \left(P_k(t, \Delta t)\right)^{\alpha_k} \cdot L_k & \text{POWER model} \\ \left[\beta_k^{\text{FAVAD}} + M_k \cdot P_k(t, \Delta t)\right] \cdot \sqrt{P_k(t, \Delta t)} \cdot L_k & \text{FAVAD model} \end{cases} \quad (2)$$

where $d_{k-leak}$ represents mean pipe outflows due to leakages during $\Delta t$ modelled using the mean pressures at pipe level, $P_k$, $L_k$ is the length of the $k$th pipe and $\beta_k^{\text{Power}}$ and $(\beta_k^{\text{FAVAD}} + M_k)$ are the deterioration factors, i.e. $d_{k-leak}$ for unit pressure. $M_k$ refers to the pipe material, which may influence the enlargement of orifices due to pressure (Van Zyl et al., 2014; Van Zyl et al., 2017), while $\alpha_k$ is an empirical parameter that depends on pipes materials and pressure and typically requires calibration or is assumed as prior.

The AMSI indicator, previously introduced, is linked to advanced hydraulic analysis as it represents a scaled $\beta_k^{\text{Power}}$ and $(\beta_k^{\text{FAVAD}} + M_k)$ (Giustolisi et al., 2024).

Assuming the Power model (Giustolisi et al., 2024), without impairing the generality of the presentation, the daily background leakage outflow for unit length in m$^3$/day/km, considering the operative cycle of $N$ steady states $\Delta t$, can be written as follows:

$$D_{k-leak} = \frac{86,400}{L_k/1,000} \cdot N^{-1} \cdot \sum_{t=1}^{N} d_{k-leak}(t, \Delta t) = 8.64 \cdot 10^7 \cdot \beta_k^{\text{Power}} \cdot \left[N^{-1} \cdot \sum_{t=1}^{N} P_k^{\alpha_k}(t, \Delta t)\right] \quad (3)$$

where $D_{k-leak}$ is the density of diffuse leakages at pipe level. The index AMSI is defined at pipe level using the scenario pressure, $P_{k-ref}$ (mean pressure in the operative cycle) and calculating $\alpha_{k-ref}$ as reported in Giustolisi (Giustolisi et al., 2024), as follows:

$$\text{AMSI}_k = \frac{D_{k-leak}}{P_{k-ref}^{\alpha_{k-ref}}} = 8.64 \cdot 10^7 \cdot \beta_k^{\text{Power}} \quad (4)$$

Note that AMSI is the deterioration factor, i.e. the density of diffuse leakages for unit pressure.

The phenomenological twin also supports the analysis at DMA level, which accounts for topology changes due to assumptions of closed gates at the boundary of DMAs during the hydraulic analysis. This way, it is possible to scale AMSI at DMA level, as follows:

$$\text{AMSI}_{\text{DMA}} = \frac{D_{\text{DMA-leak}}}{\left(P_{\text{DMA-ref}}\right)^{\alpha_{\text{DMA-ref}}}} = 8.64 \cdot 10^7 \cdot \beta_{\text{DMA-leak}}^{\text{Power}} \tag{5}$$

where $P_{\text{DMA-ref}}$ and $\alpha_{\text{DMA-ref}}$ are the mean values of $P_{k\text{-ref}}$ and $\alpha_{k\text{-ref}}$ of pipes in each DMA, which are weighted by their lengths, while $D_{\text{DMA-leak}}$ [m³/day/km] is the density or linear water loss indicator of the DMAs.

Therefore, AMSI$_{\text{DMA}}$ is the effect of pipe deterioration at DMA level due to pressure status, i.e. a *leakage effect* due to the pressure variability among pipes combined with their specific deterioration.

This means that AMSI is not simply related to the summation of the outflow coefficients of the diffuse punctual leakages in DMA as if it is assumed a constant pressure status. Therefore, if a punctual leakage evolves, AMSI$_k$ increases because $\beta_k^{\text{Power}}$, consequently also AMSI$_{\text{DMA}}$.

The next paragraph exploits the use of the *random database*, generated through advanced hydraulic analysis, to develop a strategy for pressure sampling design, which is based on the leakage detection performance, *error-free*, i.e., setting null measurement errors, and engineering judgment.

## 5. Strategy for pressure sampling design

To demonstrate and discuss the strategy for pressure sampling design, a real small-sized WDN is here used. It is the hydraulic system serving the municipality of Roccaforzata-Monteparano, in southern Italy, supplying about 4,000 inhabitants with 1,619 consumer meters.

The network consists of 373 pipes and 311 nodes with a total length of about 27 km, fed by two reservoirs; the pressure control is achieved with two valves. The system is divided into five DMAs as shown in Figure 3.

The hydraulic model was calibrated using real data provided by the water utility, Acquedotto Pugliese, so the relevant parameters of normal conditions, i.e., $D_{\text{DMA-leak}}$, $P_{\text{DMA-ref}}$ and AMSI$_{\text{DMA-leak}}$, for each DMA, are known. This allows to calculate the variation of AMSI$_{\text{DMA-leak}}$ due to the anomaly, assumed an orifice at pipe level, and the *scenario* and *random* databases.

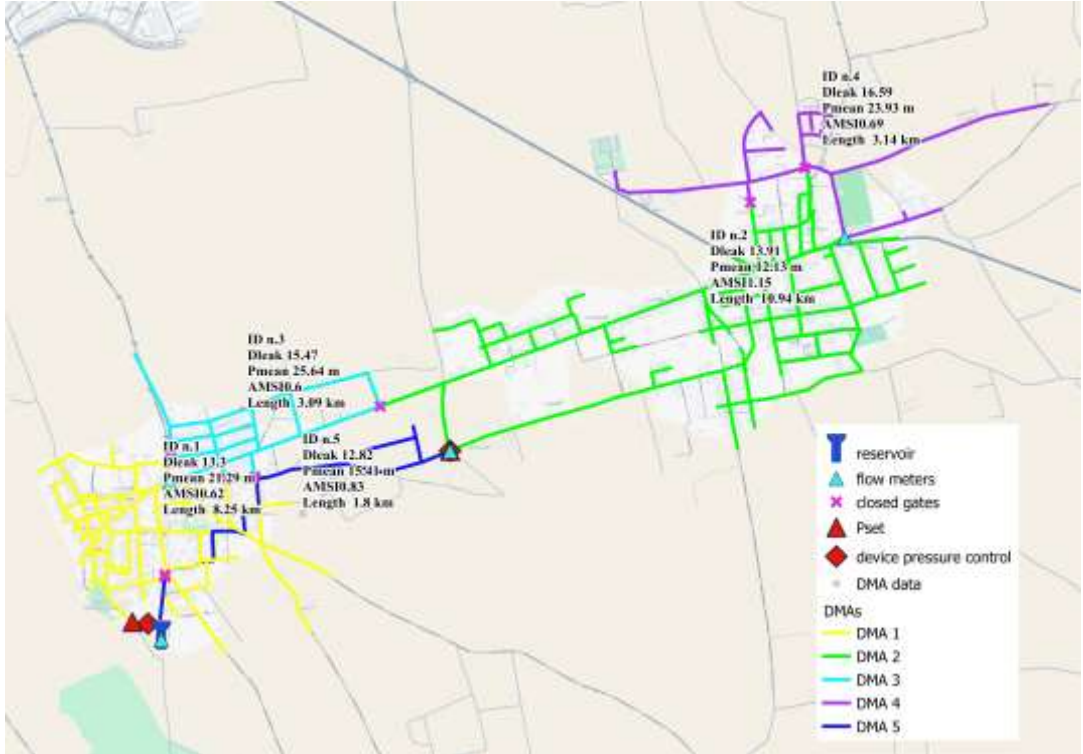

Figure 3. WDN of Roccaforzata-Monteparano.

The *scenario database* was generated by sampling ten orifice values in the range from 0.5 to 1 cm and ten different levels of consumer consumption demand for each pipe, thus generating a variety of boundary conditions across the network. Note that the maximum value of the orifice (1 cm) was assumed considering the hydraulic capacity of the system, i.e., higher sizes generate punctual leakage outflow whose detection becomes trivial.

The *random database* was here used for pressure meters sampling design. The number of events in the *random database* was set 10,000 that is a number of samples statistically robust for the WDN size.

As previously reported, the districtualization provides a structured framework for the preliminary location of the pressure meters at the boundary of each DMA, named *boundary DMA* configuration, specifically at flow meter and closed gate locations. This configuration also ensures a baseline for computing $P_{DMA\text{-}ref}$ and collecting pressure data to support punctual leakage pre-localization, after DMA identification, by determining the sequence of pipes to inspect.

The driver indicators of the sampling design are the percentage of pipes affected by a punctual leakage correctly pre-localized as first in the inspection sequence (*always predicted*) and never pre-localized as first (*never predicted*); the measurement errors are assumed null (*error-free* condition).

The ratio of the reported indicators is related to the fact that pressure meters are in few nodal positions and, consequently, they represent a sampling of the pressure network status.

Therefore, for hydraulic/topological reasons, the leakage events in two pipes can correspond to very similar sampled pressure drops in the *scenario database*. This fact can mask the pre-localization of one with respect to the other.

The pressure sampling design strategy of the *boundary DMA* configuration is refined adding pressure meters internally to the DMAs, *internal DMA* configuration, and to the ending nodes of peripheral pipes, *peripheral* configuration. The refinement is supported by expert engineering judgment and the maximization of the *always predicted* indicator and the minimization of the *never predicted* one. Expert engineering judgment involves locating the pressure meters by considering the complexity of the technical constraints and making decisions based on the added value with respect to the previous indicators.

The reason of pressure meters internal to DMA is to improve the pre-localization of internal pipes, while the reason of peripheral is their sensitivity to events.

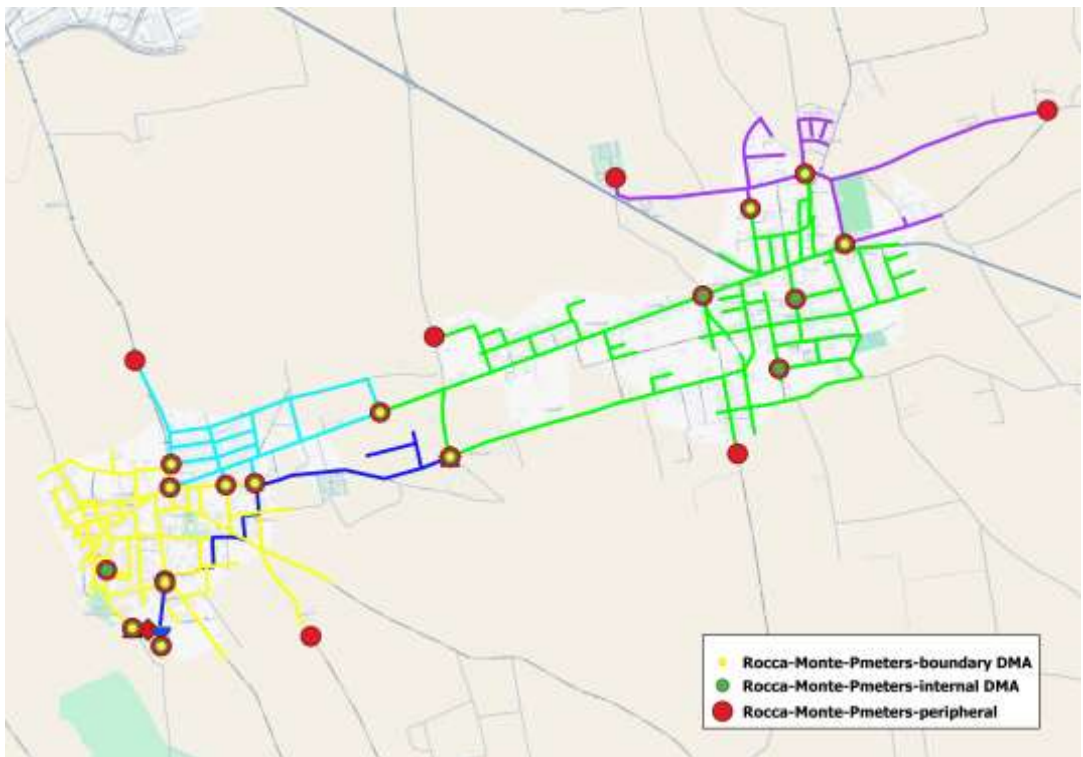

Figure 4. Pressure metering system in boundary, internal and peripheral configurations

Figure 4 reports the pressure sampling design, from the *boundary DMA* to *internal DMA* and *peripheral* configurations. Table I reports the performance of such configurations, in terms of *always predicted* and *never predicted* as first, showing the progressive improvement in performance going towards denser metering solutions. For completeness, Table I also reports additional indicators, named *predicted* and *not predicted*, which are the percentage of pipes correctly predicted or not predicted as first, at least once in the inspection sequence.

Table I Performance indicator results for boundary, internal and peripherals configurations

|  | Boundary DMA | Internal DMA | Peripheral |
|---|---|---|---|
| Always predicted | 48.53 % | 58.71 % | 60.86 % |
| Never predicted | 0.80 % | 0.54 % | 0.54 % |
| Predicted | 99.20 % | 99.46 % | 99.46 % |
| Not predicted | 51.47 % | 41.29 % | 39.14 % |
| n. pressure meters | 22 | 26 | 32 |

Note that the *not predicted* and *predicted* indicators are fully complementary to the *always predicted* and *never predicted* ones, respectively, being both the summation 100%; therefore, using the latter as indicators to be maximized or minimized, respectively, can be considered sufficient.

The *always predicted* indicator in the first row of Table I shows that the *boundary DMA* configuration, the structured one, has good performance. In fact, each event is always correctly predicted as first in the inspection sequence for the 48.53 % of pipes, while for the remaining 51.47 %, it is at least once not predicted as first (*not predicted*). Figure 5 reports the corresponding pipes in the WDN.

The *never predicted* indicator in the second row of Table I also shows good performance of the *boundary DMA* configuration. In fact, each event is never correctly predicted as first for the 0.80 % of pipes, while for the remaining 99.20 %, it is correctly pre-localized as first at least once (*predicted*). Figure 6 reports the corresponding pipes of the WDN.

Note that the indicators of the *boundary DMA* configuration, i.e., the performance, are very good due to the small size and low complexity of the WDN.

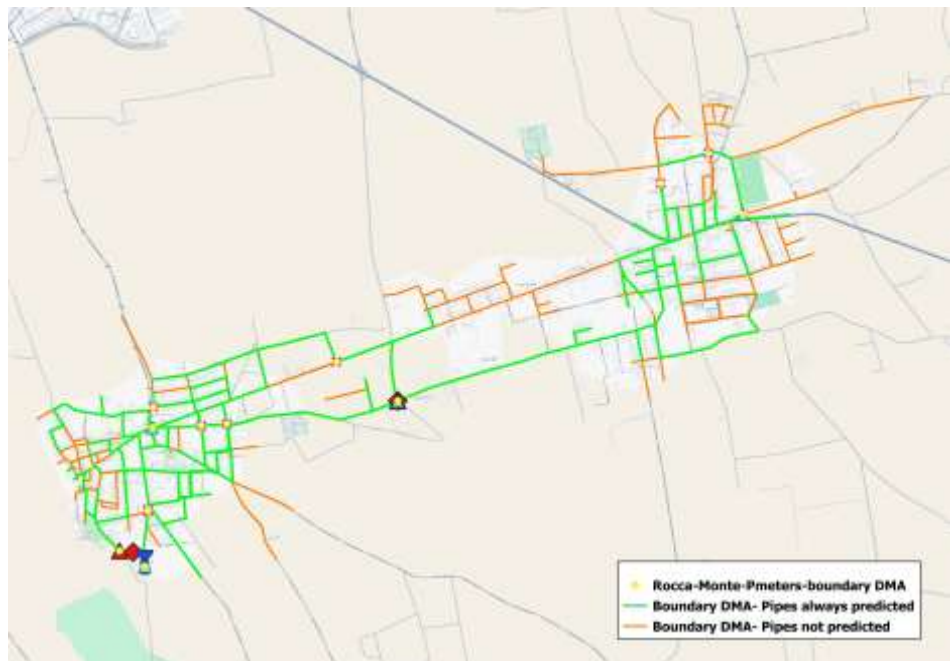

Figure 5. Pipes always predicted and not predicted in boundary configuration

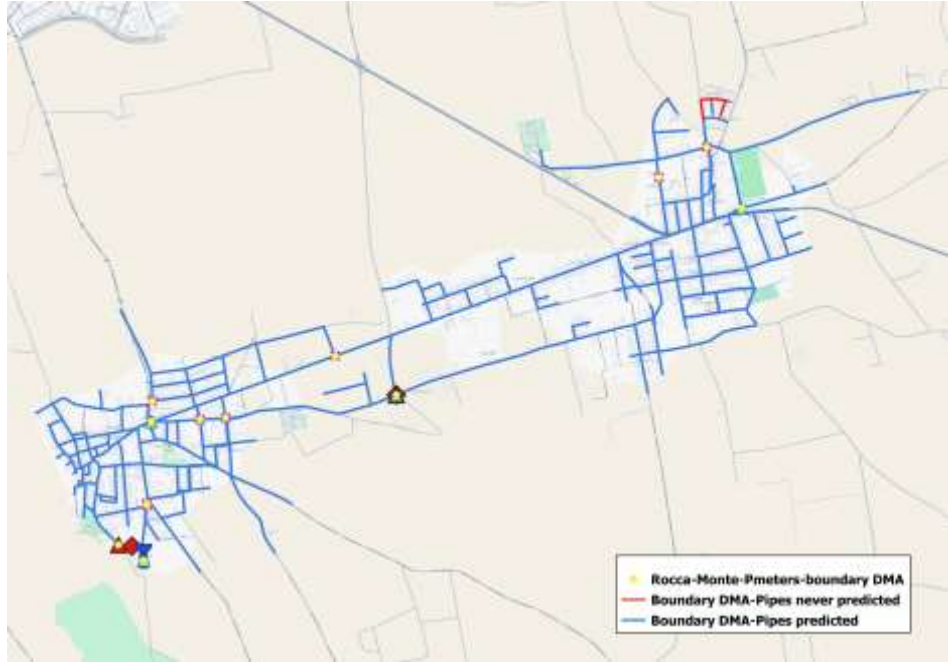

Figure 6. Pipes never predicted and predicted in boundary configuration

The second column of Table I shows the performance of *internal DMA* configuration that improves with respect to the initial configuration using four new meters. Each event is always correctly predicted as first in the inspection sequence for the 58.71 % of pipes and never correctly predicted as first for the 0.54 %. Finally, the third column of Table I shows the performance of *peripheral* configuration that improves with respect to the *internal DMA* using six new meters. Each event is always correctly predicted as first in the inspection sequence for the 60.86% of pipes and never correctly predicted as first for the 0.54 %. It worth to note that the reported strategy is instrumental to support the pressure sampling design using rational indicators (hydraulic and topology driven) and engineering judgment. Finally, note that the study of pressure sampling design performance, varying the number of pressure meters or with respect to other strategies, are beyond the scope or focus of the present work.

6. **Case study**

To demonstrate and discuss the innovative two phases strategy for identification and pre-localization of unreported punctual leakages, a large-sized WDN is here used. It is the hydraulic system serving the municipality of San Marzano, in southern Italy, supplying about 8,800 inhabitants with 3,286 consumer meters.

The network consists of 853 pipes and 684 nodes with a total length of about 51.5 km, fed by a reservoir through two main pipelines; the pressure control is achieved with two valves. The system is divided into nine DMAs as shown in Figure 7.

The hydraulic model was calibrated using real data provided by the water utility, Acquedotto Pugliese, so the relevant parameters of normal conditions, i.e., $D_{DMA\text{-}leak}$, $P_{DMA\text{-}ref}$ and $AMSI_{DMA\text{-}leak}$, for each DMA, are known. This allows to calculate the variation of $AMSI_{DMA\text{-}leak}$ due to the anomaly, assumed an orifice at pipe level, and the *scenario* and *random* databases.

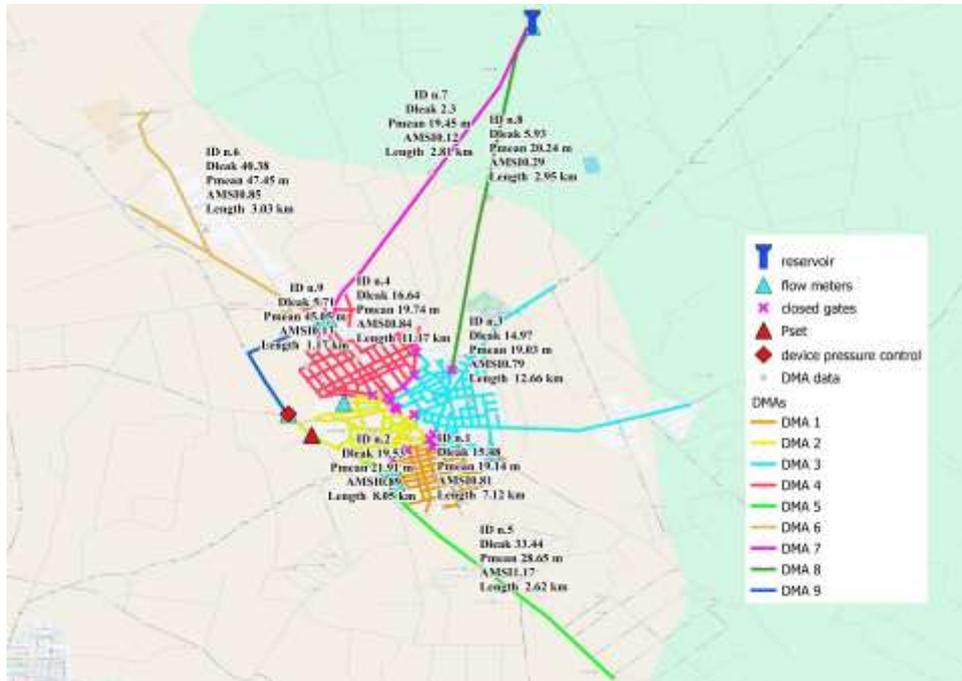

Figure 7. Water distribution network of San Marzano and DMAs

The *scenario database* was generated by sampling ten orifice values in the range from 0.5 to 2 cm (considering the hydraulic capacity of this system) for each pipe, and ten different levels of consumer consumption demand.

The *random database*, set to 10,000 number of events, was here used to test the strategy of identification and pre-localization of unreported punctual leakages, i.e., considering *error-free* condition, *threshold for AMSI variation* and *measurement errors*.

Starting from *error-free* condition, the strategy of pressure sampling design, discussed in the previous paragraph, was applied. Figure 8 reports the pressure metering system composed of sixty-nine meters according to the configuration defined as *peripheral*, which includes forty-six meters at the boundary of DMAs, thirteen meters inside the DMA and further ten in peripheral positions, generally at the ends of long branches.

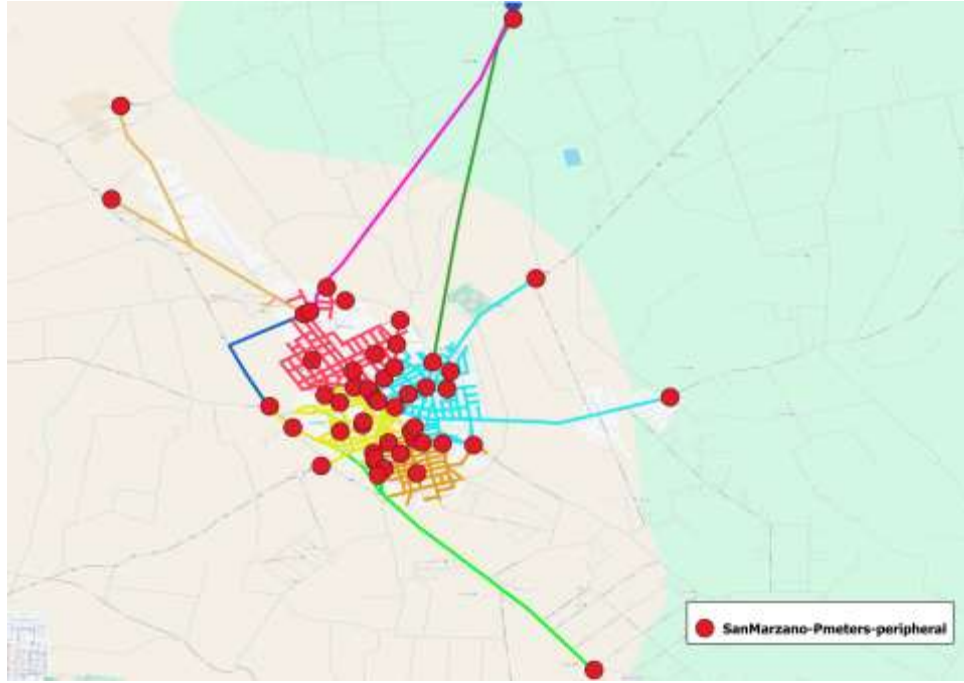

Figure 8. Pressure metering system in *peripheral* configuration

To analyse the results of the strategy, four statistical indicators were introduced:
- *true prediction*, i.e., the percentage true prediction of the pipe where the event occurs.
- *average prediction*, i.e., the average position of the pipe where the event occurs in the inspection sequence (see previous paragraph).
- *length to inspect*, i.e., the average length to be inspected to find the pipe where the event occurs.
- *percentage to inspect*, i.e., the previous indicator in percentage of the total network length.

The previous indicators were evaluated when the DMAs are identified (true positive) considering four outflow classes of the punctual leakages events: from 0.19 to 1 L/s comprised (*small outflows*), from 1 to 2 L/s comprised (*medium-small outflows*), from 2 to 3 L/s comprised (*medium-high outflow*), and higher than 3 L/s (*high outflows*). In addition, the indicators were computed considering all the outflows (*any flow*). Note that minimum outflow (0.19 L/s) represents the lowest in the *random database*.

These four outflow classes are not statistically based; rather, they represent a functional classification aimed at assessing the identifiability of the events within the network in relation to its topology, the configuration of DMAs, and the hydraulic capacity of the system. This is an important feature, for characterizing the behaviour of the specific hydraulic system. For instance, since the sampling of the orifice size is homogeneous, the prevalence of events in *medium-high* and *high outflow* classes suggests that the system operates under high pressure conditions, which generate significant outflows even from small orifice. Conversely, the prevalence of events in *small* and *medium-small outflow* classes suggests that the system operates under low pressure conditions, where even a large orifice results in a small outflow.

At first, the results of the analysis in *error-free* condition are described.

Figure 9 reports the histogram of the *true prediction* indicator. As previously reported, this histogram is not intended to provide statistical information but aims to be informative about the characteristics of the WDN *versus* leakage detection. Note that the prevalence of events belongs to the *small* and *medium-small outflow* classes., suggesting a low hydraulic capacity of the system. An increasing trend line would be expected, i.e., as the outflow increases, the pre-localization improves. However, this does not occur as the events grouped in each class are not homogeneous. In fact, the events occur in different DMAs, with different number of pipes, lengths and topological complexity.

The *any flow* class is almost 68%, reminding that the *error-free* condition was assumed. For this reason, the indicators are quite good and the correct identification of DMAs through AMSI always occurred, i.e., *100% detected*.

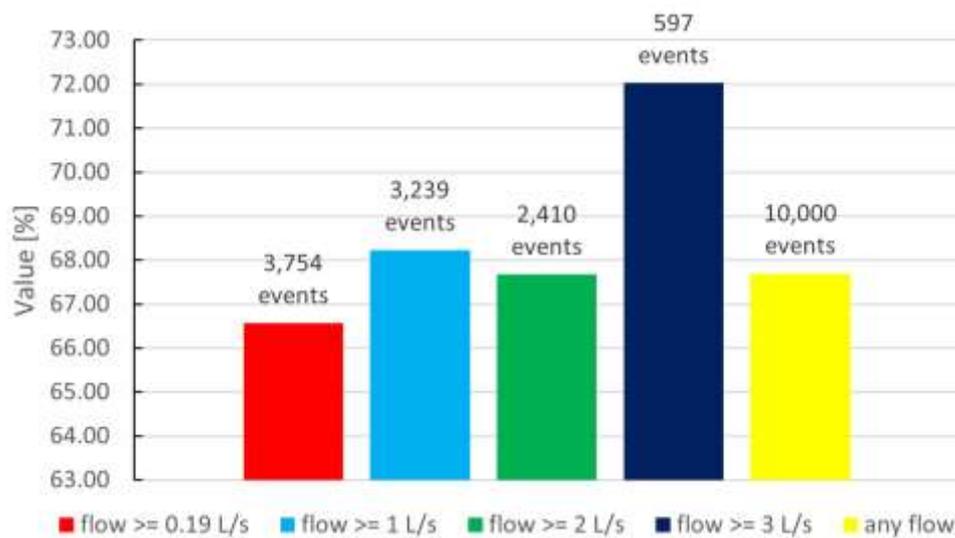

Figure 9. True prediction (100% detected)

Figure 10 reports the histogram of the *average prediction* indicator. Note that a *true prediction* value of 100% in an outflow class in Figure 9 (denoting that all punctual leakages in that class were perfectly predicted) results in an average prediction value of 1 in Figure 10. A decreasing trend line would be expected, for the same reasons stated above. However, the expected trend is masked by the non-homogeneous nature of events grouped in each class, as previously discussed. In fact, this indicator is biased by the non-constant number of pipes through DMA i.e., the maximum number to inspect is variable.

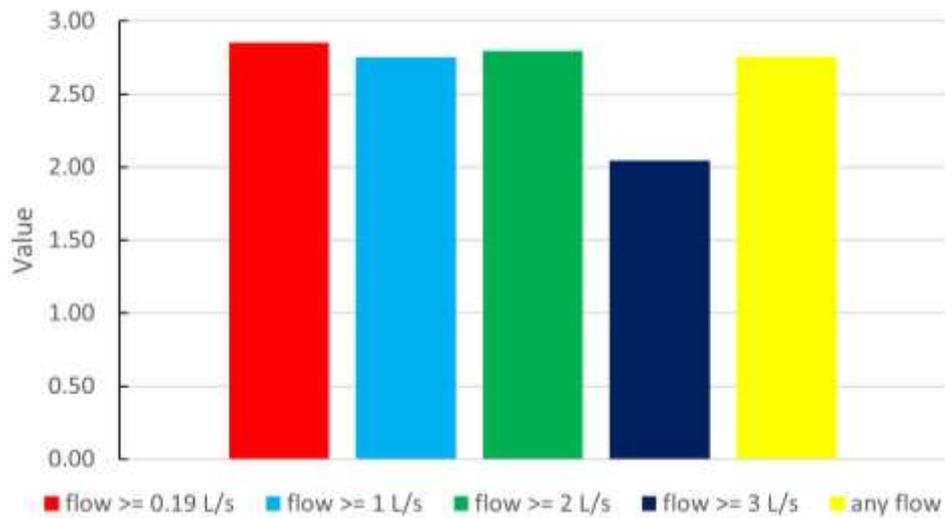

Figure 10. Average prediction

Figure 11 reports the histogram of the *length to inspect* indicator. A decreasing trend line would be expected, for the same reasons stated above. However, also in this case the indicator is biased; in addition to the non-constant number of pipes through DMAs, the average pipe length and the total length of DMAs are variable. This topological effect also explains the higher inspection lengths associated with high-flow class (Figure 11). Even if the correct pipe is ranked as first in the inspection sequence, the entire length of that pipe must be inspected, and for long pipes, the inspection remains considerable.

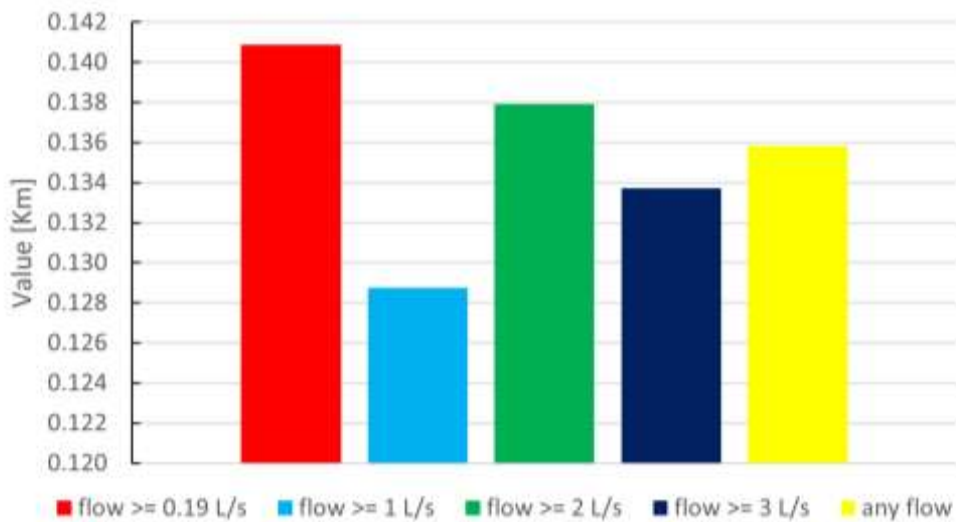

Figure 11. Length to inspect

The histogram in Figure 11 is helpful and effective for water utilities that can easily read a relevant parameter, the length to inspect, for planning their field activities. The same parameter could be calculated also at DMA level.

Figure 12 reports the histogram of the *percentage to inspect* indicator; it shares the same properties with *length to inspect*.

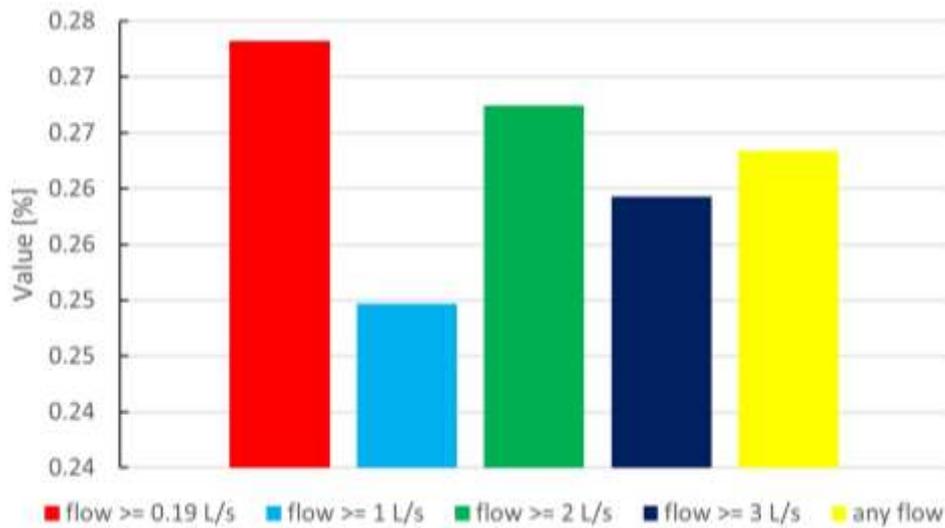

Figure 12. Percentage to inspect

However, the *percentage to inspect* indicator emphasizes the role of the number of DMAs. In absence of DMAs, a random identification of a leakage would require inspecting, on average, the 50% of the entire network length. Dividing the network into *n* DMAs, the random inspection would require inspecting the 50/*n* % of the entire network length, where *n* is the number of DMAs. In our case, the value of the indicator is equal to 50/9 = 5,56 % if the identification of the DMAs does not fail. The fact that the percentage is quite lower than 5,56 % for each class and *any flow* in Figure 12 means that the leakage detection strategy is effective.

To complete the analysis, Table II reports the relevant characteristics for each DMA: the first three columns report the specific characteristics i.e., *Number of pipes*, *Total length*, *Average pipe length*; the fourth column reports the *Average outflow* of punctual leakages; the last four columns report *Prediction Index* and the *Average inspection length*, introduced to assess the pre-localization in *error-free* condition and with *measurement errors* (real condition), for each DMA.

Table II Relevant characteristics of DMA, Prediction Index and Average Inspection Length

|  | Number of pipes | Total length [m] | Average pipe length [m] | Average Outflow [L/s] | Prediction Index (error-free) | Average inspection length [m] (error-free) | Prediction Index (with errors) | Average inspection length [m] (with errors) |
|---|---|---|---|---|---|---|---|---|
| DMA 1 | 151 | 7,120 | 47 | 1.44 | 1.005 | 75.8 | 1.409 | 2,873.6 |
| DMA 2 | 201 | 8,050 | 40 | 1.63 | 1.002 | 55.9 | 1.306 | 2,442.8 |
| DMA 3 | 256 | 1,266 | 49 | 1.42 | 1.007 | 123.1 | 1.463 | 5,925.8 |
| DMA 4 | 218 | 11,170 | 51 | 1.40 | 1.017 | 213.4 | 1.408 | 4,414.1 |

|         | Number of pipes | Total length [m] | Average pipe length [m] | Average Outflow [L/s] | Prediction Index (error-free) | Average inspection length [m] (error-free) | Prediction Index (with errors) | Average inspection length [m] (with errors) |
|---------|-----------------|------------------|-------------------------|-----------------------|-------------------------------|---------------------------------------------|--------------------------------|---------------------------------------------|
| DMA 5   | 6               | 2,616            | 374                     | 1.74                  | 1.117                         | 573.0                                       | 1.359                          | 1,107.2                                     |
| DMA 6   | 10              | 3,027            | 275                     | 2.24                  | 1.041                         | 267.5                                       | 1.456                          | 1,541.0                                     |
| DMA 7   | 2               | 2,814            | 703                     | 2.11                  | 1.000                         | 1,407.1                                     | 1.667                          | 2,657.6                                     |
| DMA 8   | 3               | 2,946            | 737                     | 2.04                  | 1.357                         | 1,600.5                                     | 1.486                          | 2,049.3                                     |
| DMA 9   | 2               | 1,165            | 583                     | 2.23                  | 1.000                         | 389.0                                       | 1.600                          | 1,891.03                                    |

The *Prediction Index* provides a comparable measure of the effectiveness of pre-localization across DMAs with respect to the number of pipes. It is calculated as:

$$\text{Prediction Index} = 1 + \frac{(P-1)}{(N-1)} \qquad (6)$$

where *P* is the average of positions in the inspection sequence for each event in the specific DMA, and *N* is the total *Number of pipes* in the DMA.

This index varies in a range [1, 2], where 1 represents the best pre-localization, where all pipes are always predicted as first, and 2 represents the worst pre-localization, where all pipes are always predicted as last. Note that the value 1.5 means the correct pipe is identified in the middle of the sequence on average, i.e., it corresponds to a random search in the DMA.

The *Average inspection length* provides a most operationally measure of the effectiveness of pre-localization across DMAs with respect to the *Average pipe length*. It is calculated as the average of lengths to inspect for each leakage event occurred in a specific DMA.

This analysis at DMA-level allows a more detailed evaluation, highlighting that outflow is not the only determining factor for punctual leakage pre-localization. In fact, the specific characteristics of each DMAs, i.e., *Number of pipes*, *Total length*, *Average pipe length,* impact on pre-localization and consequently, on the length to be inspected.

As explained in the third paragraph, the strategy allows to implement a *threshold* for AMSI *variation*, meaning that an anomaly corresponds to an effective leakage only when the threshold value is exceeded. Therefore, setting a *threshold* for AMSI *variation* avoids false positives, ensuring that only significant anomalies are considered from a technical perspective to help field inspections. Thus, *a threshold for AMSI variation* equal to 0.1 m was set.

Moreover, the strategy allows to implement *pressure measurement errors* i.e., errors relating to the pressure variation at meters that impact on the pre-localization phase.

Thus, *pressure measurement errors* equal to 0.5 m were considered.

Note that the system works considering the average daily status of the WDN, reducing the effect of measurement errors in the pre-localization phase.

The following figures report the results of the analysis considering both *threshold for AMSI variation* and *pressure measurement errors.* Figures 13 and 14 report the histogram of *true prediction indicator* and the histogram of *percentage to inspect* indicator, respectively.

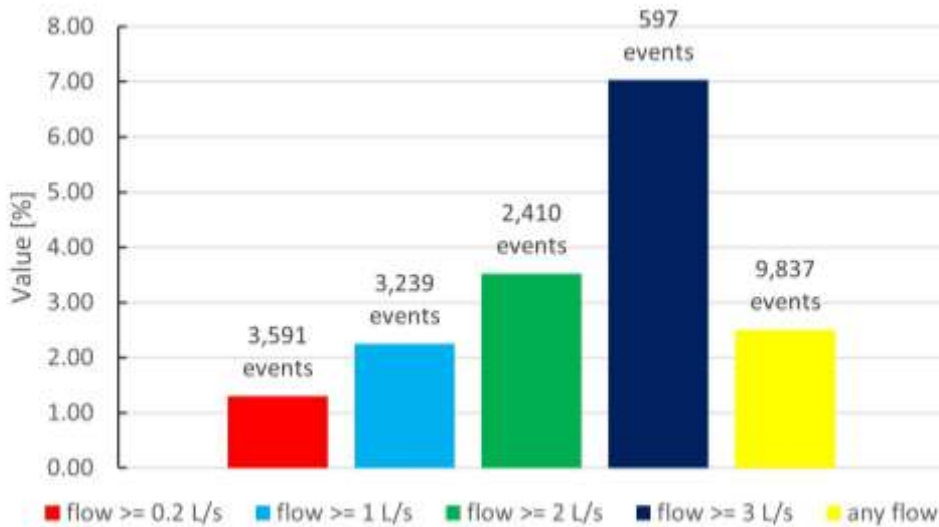

Figure 13. True prediction with threshold for AMSI variation and pressure measurement errors (98.37% detected)

With respect to *error-free* conditions (100% detected), reported in Figure 9, Figure 13 shows that the correct identification of DMAs results in 98.37 % detected. This means that some events of random database have not been associated with any DMA, as the AMSI variation does not exceed the defined threshold. In addition, Figure 13 shows an increasing trend as the outflow increases, because *pressure measurement errors* mask the effect of the heterogeneity of events in each class, i.e., non-constant number of pipes through DMAs, the average pipe length and the total length of DMAs. Furthermore, the *threshold for AMSI variation* and the *pressure measurement errors* influence the value of the percentage of true prediction, thus resulting in the decreasing of the number of pipes predicted as first. This can be also observed in Table II, where the *Prediction Index* (with errors) for each DMA gets worse by almost 40% on average.

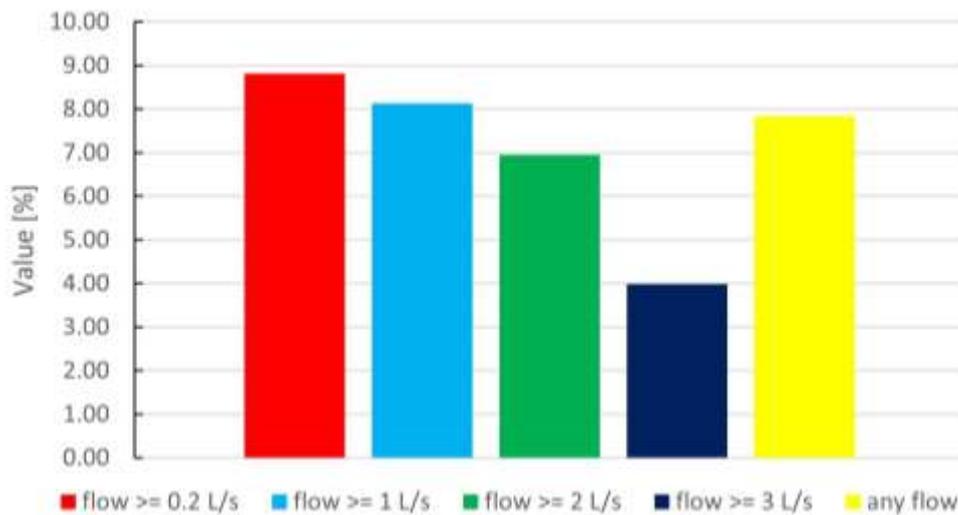

Figure 14. Percentage to inspect with threshold for AMSI variation and pressure measurement errors

With respect to *error-free* conditions, reported in Figure 12, Figure 14 shows a decreasing trend as the outflow increases, for the same reasons stated above. Furthermore, with the *threshold for AMSI variation* and the *pressure measurement errors*, the value of the percentage to inspect increase approximatively from 0.27 % to 8% for *any flow*, thus resulting in higher length to inspect to find the leakage.

This can be also observed in Table II, where the *Average inspection length* (with errors) increases by 2 km on average.

## 7. Conclusions

It is presented a novel two phases strategy *model-based* strategy for the identification and pre-localization of punctual leakages. The first phase identifies the DMAs, also more than one, having an increase of AMSI, a deterioration index, due to the evolution of the size of a punctual leakage. For this reason, the strategy is useful for DMA management and not constrained to a single anomaly. Furthermore, the second phase returns for each identified DMA a sequence of pipes to inspect which is effective for water utilities. In fact, topological features can drive the inspection of the teams of the water utility, e.g. adjacent pipes, and environmental constraints e.g. traffic, accessibility, etc., make it more efficient.

The results demonstrated that (i) the specific characteristics of each DMA and their topological complexity influenced prediction capabilities in *error-free* condition, (ii) a *threshold for AMSI variation* avoids false positives into the identification phase and (iii) the presence of *pressure measurement errors* significantly impacts on the pre-localization phase.

Future works will aim at incorporating *prior information* related to the asset (e.g., pipe age, material, diameter and connections to properties etc.) and field activities (e.g., like historical failures, pipe repairs, noise loggers, correlators, etc.) to enhance the prelocalization phase, i.e. *prior information* can be progressively part of the design of the *sequence of pipes* to inspect, enhancing leakage detection over time. Note that the reported strategy is appealing for water utilities because it is designed to be developed and customized integrating information and data, making them useful for district management, and creating value from different management activities.

**Data Availability**

The datasets analysed during the current study, i.e., data of the water distribution network, are not publicly available since they are considered sensitive by water utilities but are available from the corresponding author on reasonable request. All the data generated by analyses during this study are included in this published article.

**CRediT authorship contribution statement**

**G. Messa:** Data Curation, Investigation, Methodology, Writing – original draft, Writing – review & editing

**G. Acconciaioco:** Data Curation, Investigation, Methodology, Writing – original draft, Writing – review & editing

**S. Ripani:** Data Curation, Investigation, Methodology, Writing – original draft, Writing – review & editing

**L. Bozzelli:** Methodology, Writing – original draft, Writing – review & editing

**A. Simone:** Investigation, Methodology, Visualization, Writing – original draft, Writing – review & editing

**O. Giustolisi:** Conceptualization, Methodology, Software, Supervision, Writing – review & editing

**Funding sources**

This research did not receive any specific grant from funding agencies in the public, commercial, or not-for-profit sectors.

**Declaration of Competing Interest**

The authors declare that they have no known competing financial interests or personal relationships that could have appeared to influence the work reported in this paper.